# SUPERCONDUCTING DC AND RF PROPERTIES OF INGOT NIOBIUM[*]

Pashupati Dhakal[#], Gianluigi Ciovati, Peter Kneisel, and Ganapati Rao Myneni
Jefferson Lab, Newport News, VA 23606

*Abstract*

The thermal conductivity, DC magnetization and penetration depth of large-grain niobium hollow cylindrical rods fabricated from ingots, manufactured by CBMM subjected to chemical and heat treatment were measured. The results confirm the influence of chemical and heat-treatment processes on the superconducting properties, with no significant dependence on the impurity concentrations in the original ingots. Furthermore, RF properties, such as the surface resistance and quench field of the niobium rods were measured using a $TE_{011}$ cavity. The hollow niobium rod is the center conductor of this cavity, converting it to a coaxial cavity. The quench field is limited by the critical heat flux through the rods' cooling channel.

## INTRODUCTION

Bulk niobium (Nb) has been the material of choice for the superconducting radiofrequency (SRF) cavity used in particle accelerator because of its highest transition temperature ($T_c$) and highest lower critical field ($H_{c1} \sim 190$ mT) among the elemental type-II superconductors [1]. The overall performance of these SRF cavities is measured by its quality factor, $Q_0$ as a function of the accelerating gradient, $E_{acc}$. Higher $Q_0$ for the reduction of cryogenic loss and higher $E_{acc}$ for the use of high energy accelerators are desired. In the last four decades, much work has been done to push the performance of SRF cavity to its theoretical limits of the accelerating gradient ($\sim 55$ MV m$^{-1}$). One of the issues towards achieving those limits is the occurrence of a sharp increase of the radio frequency (RF) losses when the peak magnetic field, $B_p$, reaches about 90 mT, consequently limiting the operational accelerating gradient of SRF cavities [2]. This phenomenon is referred to as "high field Q-slope" or "Q-drop". Besides the Q-drop, several factors are limiting the high gradient and $Q_0$, such as, field emission, multipacting, residual resistance and thermal instabilities. Efforts have been made to overcome these problems using, buffer chemical polishing (BCP), electropolishing (EP), centrifugal barrel polishing (CBP), high temperature (HT) treatment, low temperature baking (LTB) and (or) its combination.

In recent years, large grain ingots Nb become an alternate to the fine grain Nb for the fabrication of high performance SRF cavities. Simpler fabrication procedures, potential cost reduction, higher thermal stability at 2K, as well as reproducibility in the performance of cavities has attracted the SRF communities towards the fabrication of SRF cavities with ingot Nb [3,4]. The study of the superconducting properties of these ingots Nb is important to optimize the chemical and heat treatment procedure during the fabrication of SRF cavities. Recently [5-7], the DC and low frequency magnetic and thermal properties of large-grain ingot niobium samples subjected to different chemical and heat treatment were reported. In this contribution, we extend the similar study to the cylindrical hollow rods of larger diameter (12 mm compared to previously measured 6 mm samples) fabricated from new niobium ingots, manufactured by CBMM (Companhia Brasileira de Metalurgia e Mineração), Brazil, subjected to the chemical and heat treatment used in cavity fabrication. Furthermore, RF properties such as the surface resistance and quench field of the niobium rods were measured using a $TE_{011}$ cavity [8]. The combination of these measurement techniques may help in finding correlations between superconducting parameters which can be measured more easily on samples and their RF properties, which are directly related to SRF cavity performance. So far, there has been limited availability in the SRF community of systems capable of measuring the RF surface resistance of superconducting samples at peak magnetic field values ($B_p$) greater than about 40 mT [9].

## EMPERIMENTAL SETUP

The system used to measure the thermal conductivity, DC magnetization, the penetration depth and surface pinning characteristic is described in Ref. [5]. As described in Ref. [8], the RF tests of the Nb pill-box cavity which was going to be used to measure the RF properties of the Nb rods showed very strong multipacting at $B_p \sim 20$ mT. This problem was mitigated by cutting the bottom plate of the pill-box and replacing it with a grooved plate designed to increase the frequency separation between the operating mode ($TE_{011}$) and the neighbouring $TM_{111}$ mode from the initial 7 MHz to about 32 MHz. The new bottom plate was machined from ingot Nb and is sealed to the rest of the cavity with indium wire.

It was found that the next limitation of the system to allow achieving high $B_p$-values on the surface of the rod inserted in the center of the cavity was given by the critical heat flux for He-II, because the long cooling channel in the hollow rod was only 2 mm in diameter.

This prompted the decision to increase the samples' diameter to 12 mm, corresponding to the maximum opening which could be machined on the top plate of the pill-box cavity. The diameter of the cooling channel was therefore increased by a factor of 4, as the rods' wall thickness is 2 mm.

Four different samples named F, G, H, and I from different ingots with different impurities concentrations (Table 1) were provided by CBMM, Brazil. The samples were machined to a diameter of 12 mm and 120 mm length. As mentioned above, the samples have an 8 mm diameter channel in the center (axially becoming a hollow cylinder with one end closed). These samples were subjected to chemical etching by BCP as well as heat treatment in UHV furnace. The summary of the results will be presented in this contribution.

Table1. Contents ppm (per weight) of the main interstitial impurities from the different Nb ingots and RRR obtained from the samples' thermal conductivity at 4.2 K measured as received.

| Samples | Ta | H | O | N | C | RRR |
|---|---|---|---|---|---|---|
| F | 1330 | <1 | <6 | <3 | <30 | 226±10 |
| G | 1375 | 7.1 | <6 | <3 | <30 | 197±8 |
| H | 704 | <1 | <6 | <3 | <30 | 240±9 |
| I | 708 | 5.4 | <6 | <3 | <30 | 224±8 |

## RESULTS AND DISCUSSIONS

*Thermal Conductivity*

The thermal conductivity of ingot niobium is measured using Fourier's law where the constant power is supplied to the one end of the sample (source) and other end is in thermal contact with helium bath (sink). The temperature dependence of the thermal conductivity of sample–I is shown in Fig. 1. The as machined sample didn't show any enhancement of thermal conductivity around 2K, however an enhancement of $\kappa_{2K}$ by a factor of ~3 was observed after the 800°C heat treatment for 3 hours followed by 140 °C baking for 3 hours, consistence with the results shown in Ref. [10]. As mentioned, the cavity made of large grain Nb operating around 2K may have better thermal stability due to the enhancement of thermal conductivity around 2 K. The study by Mondal *et al.* [5] showed that the phonon peak can be suppressed if the magnetic vortices were trapped in the sample acting as the scattering centres for phonons. Thus, any trapped magnetic vortices or lattice defects can suppress the phonon peak drastically.

The conduction of heat in metal is mostly dominated by the electronic conduction over the phonon contribution. However, in superconductor the electronic contribution decreases due to the reduced number of electrons, where electrons form the Cooper pairs, which do not contribute to heat conduction. At low temperature, the phonon contribution plays a significant role to the conduction of heat. The total thermal conductivity of a superconductor is the sum of the electronic conduction due to the unpaired electrons and lattice thermal conductivity as [11]

$$\kappa(T) = \kappa_e(T) + \kappa_l(T) = R(y)\left[\frac{\rho}{LT} + AT^2\right]^{-1}$$

$$+ \left[\frac{1}{DT^2 e^y} + \frac{1}{BlT^3}\right]^{-1} \quad (1)$$

where $y = \Delta(T)/K_BT$, $\Delta(T)$ being the BSC superconducting gap at temperature $T$, $R(y)$ is the ratio of the electronic contribution of thermal conductivity in the superconducting to the normal state. Also, $\rho$ is residual resistivity, $L$ is Lorentz number ($=2.45\times10^{-8}$ WK$^{-2}$), $A$ is the coefficient of momentum exchange with the lattice vibrations, $D$ is related to the phonon scattering by electron, $B$ is related to the scattering of phonon at crystal boundaries and $l$ is phonon mean free path. The parameters $\rho$ and $A$ can be obtained by driving the superconductor in the normal state by the application of magnetic field and measuring the thermal conductivity. The experimental data were fitted using Eq. (1) for the case where the sample was BCP treated followed by heat treatments as shown in Fig. 1, with $\Delta(0)/K_BT_c=1.76$, yielding $D =6.5\times10^{-3}$ m$^{-1}$WK$^3$, $B = 1.96\times10^3$ Wm$^{-2}$K$^{-4}$, where the mean free path of the sample is taken as the thickness of the cylinder, $l = 2\times10^{-3}$ m. These values are comparable to the values calculated in Ref. [11].

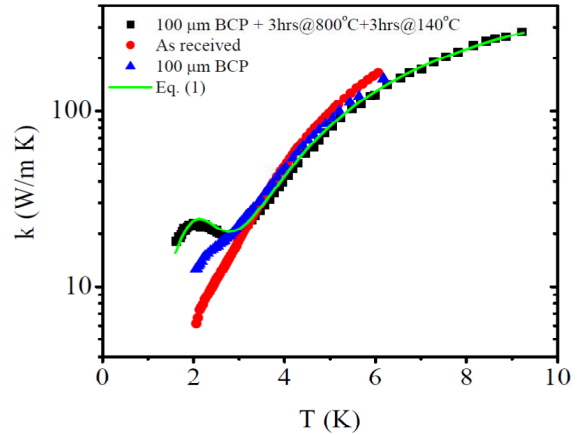

Figure 1: The temperature dependence of thermal conductivity of sample-I after various treatments. The solid line is the fit using Eq. (1) with the parameters described in text.

The thermal conductivity data can be used to estimate the residual resistivity ratio (RRR) as given by $RRR \cong 4 \kappa_{4.2K}$. The thermal conductivity of the samples measured in as received condition is shown in Table 1. The RRR values are fairly close irrespective of the impurities concentrations. The sample-G with highest tantalum and hydrogen concentrations has the lowest RRR values, whereas the sample-H has highest RRR values. It had been shown that the effect of interstitial impurities such as nitrogen, carbon and oxygen have about 20 times higher than the tantalum concentrations [12].

*DC Magnetization*

The DC magnetization measurement was carried out using single-coil magnetometer as described in Ref. [13]. The magnetic field is swept linearly at a rate of 1.3 mT/sec and the induced voltage in the pickup coil is recorded using Keithley–2182 nanovoltmeter. The magnetization $(M)$ as a function of the applied field $(H_e)$ is calculated by using the expression [10]

$$M(H_e) = \frac{1}{1-N_d} \int_0^{H_e} \frac{V(H)-V_n}{V_s-V_n} dH \qquad (2)$$

where $N_d$ (~0.01) is the demagnetization factor, $V_s$ and $V_n$ is the induced voltage in superconducting and normal state respectively.

Figure 2 shows the magnetization measurements carried out in the temperature range 2-8K of sample–I after the 100 μm BCP chemical treatments followed by 800 °C HT for 3 hours and 140 °C low LTB for 3 hours. The external magnetic field was ramped up above $H_{c2}$ and ramped down to zero. After each measurement, the sample was warmed up above the transition temperature and cooled down in zero field. The hysteresis (irreversible magnetization) in magnetization curve is observed for all temperature ranges. Several flux jumps are observed in the magnetization data as shown in Fig. 2. In case of pristine, as machined sample, the flux starts to enter the sample at 2K at an applied field ~0.4T and suddenly penetrates into the sample as seen by the sudden jump in magnetization suggesting the strong surface barrier. After surface and heat treatments both the hysteresis area and the first flux penetration field ($H_{ffp}$) are reduced, indicating the reduction of the surface barrier by these processes.

The $H_{ffp}$ ($H_{c1}$ for reversible magnetization) is extracted as the value of the applied field at which the magnetization curve deviates from the perfect diamagnetism, i.e., the position at which the magnetization curves deviate from the straight line as the external magnetic field is ramped up from zero to $H_e$. The temperature dependence of the $H_{ffp}$ and upper critical field $H_{c2}$ (extracted from the criteria $M(H_{c2}) \sim 0$, or the crossover from diamagnetic to paramagnetic state) is plotted as shown in lower inset of Fig. (2). We fitted these data using the Werthamer-Helfand-Hohenberg (WHH) formula [14] which seem to reproduce the observed $H_{ffp}$ and $H_{c2}$ temperature dependence reasonably well. We have estimated the zero temperature $H_{ffp}$ and $H_{c2}$ to be 0.191 T and 0.402 T respectively. The estimated value of $H_{ffp}$ is slightly higher than the reported value of $H_{c1}$ for good quality single crystal Nb [15].

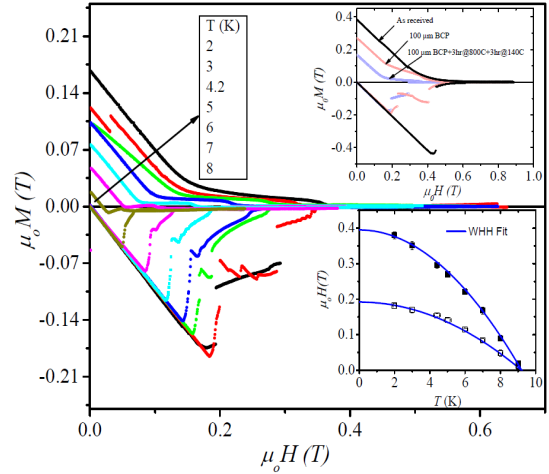

Figure 2: The magnetization measurements carried out in the temperature range 2–8 K of sample–I after BCP and heat treatment. The upper inset shows the magnetization data at 2 K of sample–I measured as received, after the 100 μm BCP chemical treatments and BCP followed by 800 °C HT for 3 hours and 140 °C low LTB for 3 hours. The lower inset shows the temperature dependence of $H_{ffp}$ and $H_{c2}$ and fit using WHH formula [14] for the data shown in the main plot.

*Penetration Depth and Surface Superconductivity*

The temperature and field dependence of penetration depth were measured by the Schawlow-Devlin method [16]. The pickup coil is connected as a part of LC oscillator and the change in resonant frequency is measured as a function of applied magnetic field and temperature. The frequency used in these experiment is ~300 kHz, with the stability of the oscillator of about $\Delta f/f \sim 10^{-5}$, which allows us sampling of ~ 15 μm depth from the surface of Nb rods. This method has been used previously to measure the field and temperature dependence of penetration depth of Nb [17-21]. Figure 3 shows the temperature dependence of the change in resonant frequency for sample–H subjected to the BCP surface treatment followed by HT and LTB. The sharp step in change in frequency corresponds to the superconducting to normal state transition. The transition temperature for as machined sample is 9.35±0.02 K, where as BCP and heat treatments sample has $T_c$ = 9.25±0.02 K.

Figure 4 shows the change in resonant frequency of the LC oscillator as a function of external applied DC magnetic field at temperature ranges 2–8 K for sample–H, which is proportional to the change in penetration depth [17]. This will allow us to determine the surface critical fields $H_{ffp}$, $H_{c2}$ and $H_{c3}$. When the external magnetic field is ramped up, no flux is penetrating so that the penetration remains zero. Once the external field reached $H_{ffp}$, the flux starts to penetrate and the penetration depth increases gradually. The dependence of penetration depth above $H_{ffp}$ depends on the surface barrier (shielding current) as well as the surface pinning. At $H_{c2}$, the bulk sample become stable in its normal state, whereas the surface superconductivity still exists up to the $H_{c3}$ [22].

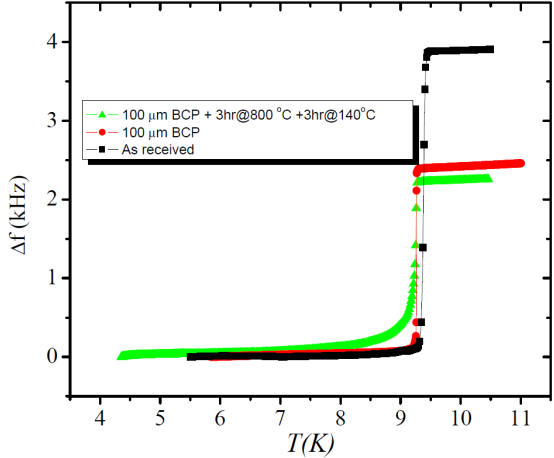

Figure 3: The change in resonant frequency of LC oscillator with temperature showing the sharp transition when the sample–H goes superconducting to normal state.

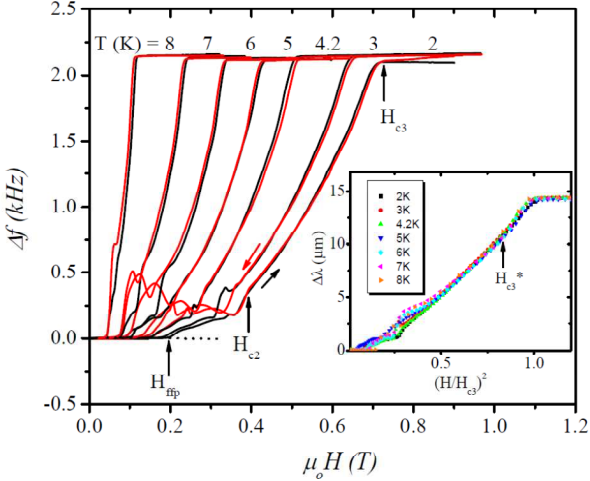

Figure 4: The change in resonant frequency of the LC oscillator with external magnetic field showing the transition at $H_{ffp}$, $H_{c2}$ and $H_{c3}$. The inset scaling of penetration depth with external magnetic field at different temperatures.

Furthermore, it is possible to measure an intermediate state between $H_{c2}$ and $H_{c3}$ at which the long-range coherent superconducting state is present at surface, called coherent surface critical field, $H_{c3}^*$ [23,24]. The scaling of the penetration depth as a function of reduced field as $(H/H_{c3})^2$ is shown in inset of Fig. 4 with the estimation (shown by arrow in inset) of coherent surface critical field ~ $(0.82 \pm 0.2)H_{c3}$, independent of the temperature. With decreasing field, the penetration depth is reversible down to $H_{c2}$ and the irreversibility observed below $H_{c2}$, depends on the surface treatment. The larger the irreversibility near $H_{ffp}$ the stronger the surface barrier. The irreversibility increases after the BCP treatment, however it decreases after heat treatments. The irreversibility can be caused by the fluxoids which are not in equilibrium with the external magnetic field with surface pinning, may be due to the oxide precipitation [19]. The irreversibility is also expected due to the slight inhomogenity of the order parameter, caused by the combination of the induced and diamagnetic screening currents, which is different in increasing and decreasing fields [25].

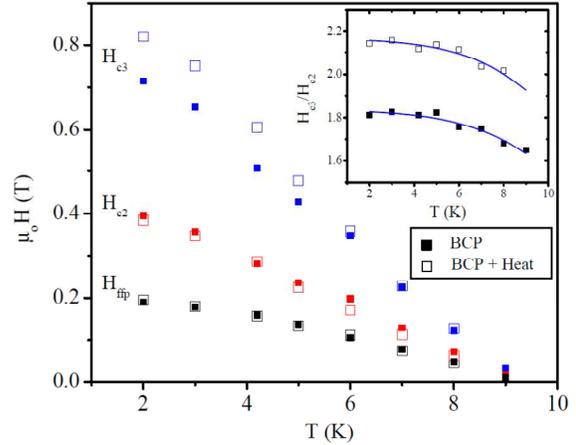

Figure 5: The temperature dependence of critical fields $H_{ffp}$, $H_{c2}$ and $H_{c3}$. The inset shows the temperature dependence of the ratio $H_{c3}/H_{c2}$. The points are experimental data whereas the straight lines are calculated using Eq. (3).

Figure 5 shows the temperature dependence of the surface critical field for sample H, subjected to the BCP and followed by heat treatment (800 °C for 3hrs and 140 °C for 3 hrs). It is seen that the $H_{ffp}$ and $H_{c2}$ stay same whereas $H_{c3}$ increases due to the heat treatment procedure as reported previously [5,6,22], yielding the increase in the ratio of $H_{c3}/H_{c2}$ as shown in the inset of Fig. 5. The existence of the surface critical field and the ratio $H_{c3}/H_{c2}$ = 1.69 near $T_c$ was predicted by Saint-James and de Gennes by solving the linearized Ginzburg–Landau equation [26]. Experimentally, the temperature dependence was observed as seen in the inset of Fig. 5. Hu and Korenman derived the temperature dependence of $H_{c3}/H_{c2}$ using the Gorkov's gap equation taking into account the nonlinearity of the microscopic theory in clean limit [27] as

$$\frac{H_{c3}}{H_{c2}} = r_{32}[1 + 0.614\,(1-t) - 0.577\,(1-1)^{3/2}$$
$$- 0.007(1-t)^2 + 0.106(1-t)^{5/2}] \quad (3)$$

where, $t = T/T_c$. The Eq. (3) fits well with the experimental data as shown in Fig. 5 inset with the value of $r_{32} = 1.61$ and 1.9 for BCP and BCP followed by heat treatment samples respectively, showing ratio increased by 18% due to the heat treatment.

## RF Measurements

Once the samples were measured for thermal conductivity, magnetization and penetration depth, they were inserted in a "pill-box" cavity forming a coaxial cavity. The cavity is tested both empty and with sample. The main electromagnetic parameters of the $TE_{011}$ mode calculated using SUPERFISH program are listed in Table 2. The maximum surface magnetic field occurs in the middle of the sample in coaxial geometry, while it is in the middle of the cylinder in the case of the empty cavity. The ratio of the peak magnetic field in these two locations is 2.22. Furthermore, the total power dissipated on the sample over the total dissipated power in the cavity per unit resistance $R_s$ is 30.4 %.

Table 2: Electromagnetic parameters of the $TE_{011}$ mode for the coaxial cavity.

| Parameters | Empty Cavity | Coaxial Cavity |
|---|---|---|
| Resonant frequency (GHz) | 3.501 | 3.856 |
| $B_p/\sqrt{U}$ (mT/J) | 62.7 | 114.2 |
| Geometric factor ($G = Q_0 R_s$) | 779.6 Ω | 532.2 Ω |

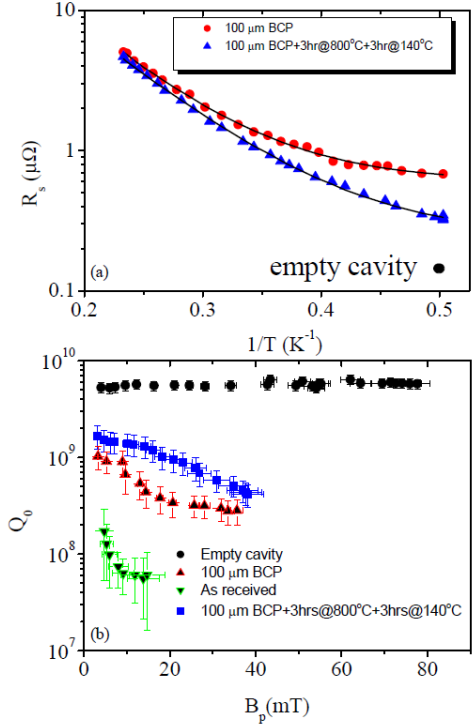

temperature dependence of the surface resistance of the cavity with sample H in the "as received" condition and after BCP and additional heat treatment. In RF field the surface resistance is the sum of the superconducting BCS resistance and the temperature independent residual resistance, i.e., $R_s = R_{BCS}+R_{res}$. For the temperature range $T<0.5T_c$, the BCS resistivity can be approximated as

$$R_{BCS} = 2\times10^{-4}\left(\frac{f}{1.5\times10^9}\right)^2 \frac{e^{-17.67/T}}{T} (\Omega) \qquad (4)$$

where, $f$ is frequency, $T$ is temperature and $T_c = 9.25$ K for niobium. We have fitted the temperature dependence of the surface resistance using the BCS codes [28], instead of the approximate Eq. (4), to extract the residual resistance. It is found to be $R_{res} = 609\pm24$ nΩ and $254\pm13$ nΩ before and after the heat treatments. Also, the BCS gap is estimated to be $\Delta/K_BT_c = 1.83\pm0.02$ and $1.73\pm0.07$ after BCP and after additional heat treatment, respectively. This shows the reduction of surface resistance due to the heat treatment and hence the increase in quality factor as shown in Fig. 6 (b). As shown in Fig. 6(b) the empty cavity reached the peak magnetic field of $B_p = 78$ mT with $Q_0 = 5.7\times10^9$, limited by quench. However, in coaxial cavity the maximum peak field was limited to about 40 mT due to the critical heat flux through the cooling channel of the sample.

The power dissipated on the sample of surface area $A_s$ is given by

$$P_{diss} = \frac{1}{2}\int_A R_s H^2 dA \qquad (5)$$

where $R_s$ is surface resistance of the sample and $H=B_p/\mu_0$, RF magnetic field. The surface resistance of the sample can be calculated from the experiment as

$$R_{s,sample} = \left(\frac{P_{cavity+sample}}{P_{cavity}}\right)\frac{R_{s,cavity}}{0.304} \qquad (6).$$

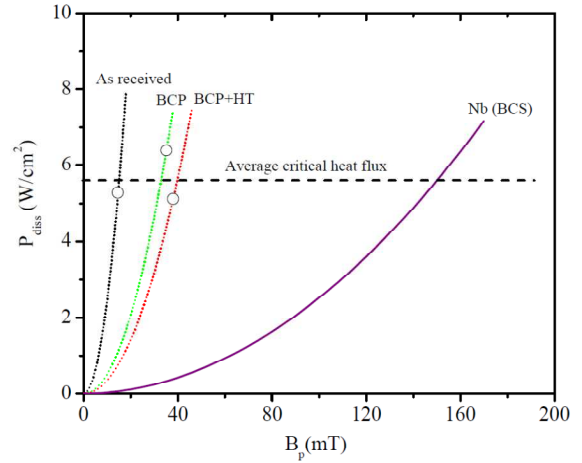

Figure 6: Summary of the RF test on the $TE0_{11}$ cavity. (a) The temperature dependence of surface resistance of $TE_{011}$ coaxial cavity with sample H after BCP and BCP followed by heat treatment at $B_p \simeq 5.5$ mT. The solid lines are the fits using BCS codes (b) The $Q_0$ of the cavity in "pill-box" and coaxial geometry with sample–H at 2K.

Figure 6 shows the summary of the RF tests on the $TE_{011}$ cavity. The cavity is tested without the sample to obtain the baseline measurement. Figure 6 (a) shows the

Figure 7: The power dissipation in the sample as a function of peak RF field for sample-H (dotted lines). Also, shown is the ideal case for Nb ($\Delta/K_BT_c=1.9$). The points are the field at which the cavity was thermally instable. The horizontal dashed line is the average critical heat flux.

Figure 7 shows the power dissipation in the sample as a function of peak magnetic field. Also shown is the ideal case for Nb only taking into account of the BCS resistance. The average heat flux through the sample's cooling channel at the highest field achieved is 5.6Wcm$^{-2}$, comparable to the critical heat flux for He-II [29]. As shown in Fig. 7, the estimated maximum $B_p$-value which can be achieved at 2 K with the present cavity setup and an ideal (zero residual resistance) niobium sample is of the order of 150 mT.

## SUMMARY

The thermal conductivity, DC magnetization, penetration depth as well as the RF properties of ingot Nb rods which are subjected to BCP chemical and heat treatments were measured. The surface and heat treatment methods show the improvement of both the DC and RF properties of ingot niobium. These can be attributed to reduced surface and bulk pinning by these processes. The heat treatments enhance the phonon peak in thermal conductivity as well as the increase in surface critical field $H_{c3}$. However, we observed the reduction in RRR values after HT, suspected due to the contamination in the furnace during HT. Despite the difference in tantalum concentrations by a factor of two in samples, the superconducting properties of the samples behave same, consistent with the previously reported results [6,30]. RF measurement on $TE_{011}$ cavity shows the reduction of surface resistance and hence the increase in quality factor due to the surface and heat treatments. The maximum peak magnetic field is limited by the critical heat flux through the niobium rods' cooling channel but it is still among the highest values achieved on samples.

Further measurements after EP, LTB and several combinations of chemical and heat treatments are planned to understand the limiting factors of cavity performance as well as the implementation of these results to new generation SRF cavities are part of future studies. Single-cell cavities from the same ingot material from which the samples were made have been fabricated and will be tested in the near future.

## ACKNOWLEDGEMENT

We would like to acknowledge P. Kushnick and M. Morrone at Jefferson Lab for the technical support.